\documentclass[a4paper,fleqn,usenatbib]{mnras}

\usepackage[T1]{fontenc}
\usepackage{ae,aecompl}

\usepackage{graphicx}	
\usepackage{amsmath}	
\usepackage{amssymb}	
\usepackage[version=3]{mhchem} 
\usepackage{siunitx}
\usepackage{natbib}

\usepackage{wasysym}
\usepackage{float}
\usepackage{dsfont}
\usepackage{comment}

\usepackage{xcolor}
\usepackage{enumitem}

\usepackage{newtxtext,newtxmath}

\newtheorem{algo}{Algorithm}

\usepackage{xcolor}

\newcommand{\1}{{\boldsymbol 1}}
\newcommand{\xx}{{\boldsymbol x}}
\newcommand{\yy}{{\boldsymbol y}}

\newcommand{\Var}{\mathrm{Var}}

\title[Gibbs model of the galaxy distribution]{Morpho-statistical characterisation of the spatial galaxy distribution through Gibbs point processes}

\author[Ll. Hurtado-Gil et al.]{Llu\'is Hurtado-Gil,$^{1,}$$^{3}$\thanks{E-mail: lluis.hurtado@edreamsodigeo.com}
Radu S. Stoica$^{2}$,
Vicent J. Mart\'inez$^{3,}$$^{4}$ \&
Pablo Arnalte-Mur$^{3,}$$^{4}$
\\
$^{1}$ eDreams ODIGEO, C/ Bail\`en 67-69, 08009, Barcelona;\\
$^{2}$ Universit\'e de Lorraine, CNRS, IECL, F-54000 Nancy, France\\
$^{3}$ Observatori Astron\`omic, Universitat de Val\`encia, C/ Catedr\'atico Jos\'e Beltran, 2, 46980 Paterna (Val\`encia), Spain\\
$^{4}$ Departament d'Astronomia i Astrof\'isica, Universitat de Val\`encia, 46100-Burjassot, Val\`encia, Spain\\
}

\date{Submitted to arXiv - 27th April 2021}

\pubyear{2021}

\begin{document}
\label{firstpage}
\pagerange{\pageref{firstpage}--\pageref{lastpage}}
\maketitle

\begin{abstract}
This paper proposes a morpho-statistical characterisation of the galaxy distribution through spatial statistical modelling based on inhomogeneous Gibbs point processes. The galaxy distribution is supposed to exhibit two components. The first one is related to the major geometrical features exhibited by the observed galaxy field, here, its corresponding filamentary pattern. The second one is related to the interactions exhibited by the galaxies. Gibbs point processes are statistical models able to integrate these two aspects in a probability density, controlled by some parameters. Several such models are fitted to real observational data via the ABC Shadow algorithm. This algorithm provides simultaneous parameter estimation and posterior based inference, hence allowing the derivation of the statistical significance of the obtained results.
\end{abstract}




\section{Introduction}\label{intro}
Galaxies tend to clump in space. Their clustering at different scales is very well encapsulated by the two-point correlation function $\xi(r)$, which measures the clustering at distance $r$ in excess $(\xi(r) > 0)$ or in defect $(\xi(r) < 0)$ compared with a Poisson distribution, for which $\xi(r) = 0$. This function was first measured on catalogues of galaxies listing only their angular positions onto the celestial sphere, but lacking distance information~\citep{tot69,pee74}. Knowing the angular correlation function, it was possible to estimate the real-space correlation function by means of the Limber formula~\citep{pee80,mar02}. When this was done, it was evident that the correlation function $\xi(r)$ followed a power-law $\xi(r)=(r/r_0)^{-\gamma}$ (with $\gamma \simeq 1.8$ and $r_0 = 5.4 \, h^{-1}$ Mpc) in a wide range of scales $0.1\leq r \leq 15 h^{-1}$ Mpc, where $h$ is the Hubble's constant in units of $100$ km s$^{-1}$ Mpc$^{-1}$. At the mid-eighties of the previous century, once the first redshift surveys were available, it was possible to measure directly the three-dimensional correlation function and its power-law behaviour was unambiguously confirmed
~\citep{dav83}. More recent studies of the correlation function of the galaxy distribution calculated on the SDSS survey ~\citep{zehavi11} or the 2dFGRS ~\citep{Hawkins:2002sg} have confirmed its power-law behaviour, although the exponent $\gamma$ and the correlation length $r_0$ depend on the luminosity cut-off of the sample and on the spectral or morphological type of the analysed survey of galaxies. The dependence of the galaxy clustering on the luminosity or on the spectral type is not only detected in spectroscopic redshift surveys but also in deep surveys built using photometric redshifts~\citep{arnalte14,hurtado16}, providing an exceptional tool to analyse the evolution of clustering with comic time. 

Understanding the galaxy distribution of cosmic web catalogues as a point process, we assume that the Universe is a representation of a stochastic process where galaxies are randomly located points in real space. Given this theoretical framework, we can make use of different statistical methods to better understand and characterise the nature of our studied point process. Some of these techniques can be found in \cite{mar02, feigelson2012modern, fruhwirth2018handbook, BaddEtAl16}.

Many well known statistical methods can be used to study the large scale structure of the Universe. One of the basic techniques already applied by Hubble on the first galaxy maps projected on a given patch of the sky was the counts-in-cells method~\citep{hubble34}. He could see that the galaxy counts in two-dimensional cells were well approximated by a lognormal distribution. For a recent analysis of the three-dimensional galaxy distribution by the counts-in-cells method see \citet{hurtado17}. Related to the counts-in-cells, we can define the $N$-probability density function of the galaxy number density fluctuations~\citep{arnalte16}. The relation between the $n$-point correlation function and the counts-in-cells has been studied in detail by \citet{white79}. Moments of the counts-in-cells are also connected with other popular statistical descriptors of the galaxy distribution related with fractals and multifractals~\citep{martinez90,borgani95,jones04}. Some other descriptors of the structure deal with continuous density fields, and therefore a first step has to be the smoothing of the discrete point process to obtain an estimation of the underlying density field by means of a kernel or filter function (typically a Gaussian filter is used). The kernel depends on the choice of a particular bandwidth. The topological genus introduced by \citet{gott86} or the more general set of functions, the so-called Minkowsky functionals, introduced by  \citet{mecke94}, allow us to characterise the global morphology of the reconstructed density field. Moreover, they can be used to define shape-finder quantities to measure the planarity of the filamentarity of the galaxy distribution~\citep{sahni98}. Other set of summary statistics are those based on the distributions of distances to the nearest neighbours, such as the void probability function or empty space function $F(r)$ \citep{maurogordato87}, or the distribution function $G(r)$ of the distances between galaxies. A particularly useful statistic is the quotient between these quantities $J(r)=(1-G(r))/(1-F(r))$ which is a good measure of the spatial pattern interaction ~\citep{lieshout1996nonparametric, kerscher99}. 

Some of these techniques, like kernel density estimators and other non parametric methods, provide valuable descriptions of the galaxy field, but might not help us to fully characterise the real nature of the structures behind the galaxy distribution \citep{mar02}. Summary statistics, like correlation functions, are capable of providing reliable and relevant information regarding isotropic structures, such as clusters and baryon acoustic oscillations \citep{1998ApJ...496..605E}, but shrink all information, losing the detail of particular structures \citep{pee80, martinez2009reliability, Meneux06vimos, Coil08deep2, Zehavi04depart}. Considering the complex dissipative processes (dynamical friction, tidal interaction, etc) involved in the clustering of galaxies at the smallest scales, it is desirable to characterise the spatial distribution with models that are sensible to these short-scale interactions between points. The Gibbs point process introduced in this paper is suited for this purpose. Interactions are modelled through an energy function. In Gibbs point processes low-energy configurations are more likely to occur than high energy ones \citep{dereudre2018}.

Solving the problem of characterising complex galaxy structures implies formulating a probabilistic model, which understands the abundance of points for every given region as a random variable depending on the environment and covariates. Given a point process, such as a galaxy catalogue, fitting a correct stochastic model of its distribution implies solving the point process problem, i.e. completely characterising the process \citep{babu2007statistical, 2007ASPC..371...22B}.

The development of point process theory has brought a whole new branch of statistical analysis of point processes, such as the galaxy distribution \citep{moller2003statistical,Lies00,chiu2013stochastic}. The well known methodologies appear integrated together with theorems already applied in different sciences with success \citep{illian2008statistical, BAD2, pfeifer1992spatial, engle1982autoregressive, diggle1990point, stoyan2000recent}. The application of these new methodologies in cosmology is a novelty with already some remarkable success \citep{StoiEtAl07,stoica2010filaments,Tempel14}. 

Point processes are spatial stochastic processes whose outcomes are random configurations of points $\xx$. Characteristics or marks can be attached to the points. The distribution of points in an observed finite domain might be controlled by a parametric probability density. This density may take into account the special locations of the observed domain and also interactions among points. Such a probability density can be written as:
\begin{equation}
p(\xx|\theta) = \frac{\exp(U(\xx|\theta))}{Z(\theta)} = \frac{\exp\left(\sum_{i=1}^{r}\theta_{i}t_i(\xx)\right)}{Z(\theta)} \text{,}
\label{gibbsPointProcess}
\end{equation}
with $U$ the energy function, $\theta=(\theta_1,\ldots,\theta_r)$ and $t(\xx) = (t_{1}(\xx),\ldots,t_{r}(\xx))$ the parameters and sufficient statistics vectors, respectively, and $Z(\theta)$ the normalising constant function. The vector $t_i,\quad i = 1,\ldots,r$ is the sufficient statistic of the considered model~\eqref{gibbsPointProcess}, since it describes all the needed information to be computed from an observed sample, in order to estimate the model parameters. The point processes characterised by probability density~\eqref{gibbsPointProcess} are called in the literature Gibbs point processes. Without loss of generality, let us assume here, that the parameter space $\Theta$ is a compact region in $\mathds{R}^{r}$ and that $p(\xx|\theta)$ is continuously differentiable with respect to $\theta$.

All these make point processes a natural candidate for modelling the galaxy distribution. Furthermore, the sufficient statistics of the model give a complete characterisation of the observed data set, while providing their morphological and statistical description. Still, the fact that the $Z(\theta)$ is not available in analytic closed form, limited the use of this tool only for the derivation  and the application of summary statistics (pair correlation function, empty space function, K-Ripley function, etc.). Recently ~\cite{StoiEtAl17} proposed a method allowing simultaneous classical likelihood and posterior based inferences, with a computational cost comparable with the one needed for simulating the model. Within this new context, the point process can be used for modelling galaxy fields, while providing statistical significance of the obtained result. Recently, Gibbs point processes have been used to study the spatial distribution of young stellar clusters and giant molecular clouds in M33~\citep{lidayi21}. The authors show how this modelling can be used to quantify the inhomogeneity of the spatial distribution of clusters and clouds within the galaxy. In a similar way, in the present paper, we show that for describing the large-scale spatial distribution of galaxies, the use of the Gibbs point process model is quite useful to relate the filamentary pattern with the proximity of the galaxies to the filaments. 

The validation of a point process model is another challenging task, specially for 3-dimensional data sets. The residual analysis methodology proposed by \cite{RSSB:RSSB519} is a technique allowing to perform this validation. New methods have been developed by the authors, adapting the residual analysis algorithms to the 3-dimensional case. With these methods we expect to detect possible disagreements between data and model as well as to understand some limits of the proposed models.

In section~\ref{data} we summarise the main properties of the data sample used in this work. In section~\ref{models} we introduce the mathematical definition of the different studied Gibbs models. In sections~\ref{methods:estim} and~\ref{choice} we describe the model estimation methods.  Results of the model application and analysis of the obtained quantities are shown in section~\ref{results}. Finally, conclusions are presented in section~\ref{conclusions}.

\section{Data}\label{data}

In \cite{Tempel14}, the authors obtained a filament catalogue using the galaxy catalogue published in \cite{Temple12}. This catalogue is based on the SDSS DR8 \citep{2011ApJS..193...29A} and corrects the Finger-of-God effects using a friends-of-friends (FoF) algorithm. 

We select a galaxy sample from this corrected catalogue together with its filamentary structure. The data set must be selected with redshift below 0.08 to guarantee its completeness. Since our selected sample has an approximate redshift of 0.05, a magnitude threshold of $M_r - 5\log h > -18$ is imposed \citep{2014A&A...566A...1T}. From the resulting volume limited region we select a cube of size $30\,h^{-1}$ Mpc where the galaxies and the detected filaments create a rich structure. 

\begin{figure}
  \centering
\includegraphics[width=\linewidth]{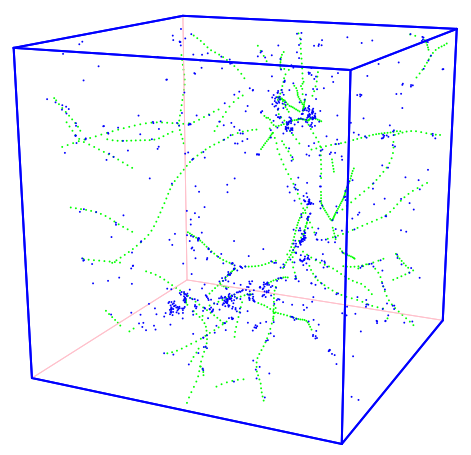}
\caption[]{Data set extracted from the SDSS catalogue as published by~\cite{Temple12} (in blue) with the filamentary structure from~\cite{Tempel14} (in green). The side of this cubic box is $30\,h^{-1}\mathrm{Mpc}$.}\label{sample1}
\end{figure}

The final selection contains 956 galaxies (number density $0.035 h^3$ Mpc$^{-3}$) and can be seen in Figure~\ref{sample1}. 
The used data set is large enough to show the clustering and filamentary patterns of the galaxy distribution. 
This sample is satisfactory to illustrate the Gibbs point process model in the context of the analysis of the morpho-statistical properties of the spatial galaxy distribution.
As the main aim of this work is to test the applicability of such models to the galaxy distribution, we do not use a larger sample which would impose more challenging computational needs.

\section{Gibbs point processes}
\label{models}
A point process is a stochastic process whose outputs are a random configuration of points $\xx = \{x_i\}_{i=1}^n$ within a given window $W$. Here, the point process is assumed locally finite (the number of points $n(\xx \cap W)$ is a finite number for a finite volume $W$) and simple ($x_i \neq x_j$ for $i \neq j$). Points might have information attached, that we call marks. The marks $M = \{m_i\}_{i=1}^n$ contain all relevant information regarding any individual point besides its spatial location.

Gibbs point processes \citep{illian2008statistical,BaddEtAl16,chiu2013stochastic} is an important class of point processes. These models understand the realisations of the process as determined by a trend or field density and an interaction function of the points. The specification of these two components leads to different processes with peculiar properties.

Maybe the best known point process is the homogeneous Poisson process. The points are spread uniformly in the observation domain, the number of points in a bounded region follows a Poisson distribution of constant intensity parameter $\rho$ and the counts of points in disjoint bounded regions are independent random variables. Due to these properties, this process is used  to model {\it completely random} point patterns, and it is commonly used as null hypothesis whenever statistical tests are implemented. Gibbs models allow the generalisation of these processes, by means of a trend function and by specifying point interaction. The trend controls the variability of the process intensity, creating areas where points appear in different abundances. If only such a trend is present, the process is called Inhomogeneous Poisson point process. The points may also interact with each other, creating repulsive or clustered point patterns.

\subsection{Definition}

Let $X$ be a finite point process in $W$. Its density is defined with respect to the unit intensity stationary Poisson process in $W$. For interacting point processes its expression may be written as 
\begin{equation}\label{pairpdf}
p(\xx) = \alpha \prod_{i=1}^{n(\xx)} b(x_i) \prod_{i<j}h(x_i,x_j).
\end{equation}

The function $b(\cdot)$ represents the trend of the process and it might be used to model external influences in the distribution or a large scale pattern. The symmetric function $h(x_i,x_j)$ represents the interaction. Here only interactions between pairs of points are considered. The normalising constant of the probability density is given by $\alpha$.

By taking the logarithm of the un-normalised right term in~\eqref{pairpdf}, the energy of a Gibbs point processes is obtained. This leads towards a very well known form of the probability densities describing this type of processes:
\begin{equation*}\label{pdfGibbs}
p(\xx) = \alpha \exp[-U(\xx)].
\end{equation*}

Whenever the interaction range is limited, this may allow the local specification of models~\eqref{pairpdf}, hence simplifying its use and expression. These processes are known in the literature as {\it Markov point processes}~\citep{Lies00}.

\subsection{Papangelou conditional intensity}

The main difficulty in using Gibbs point process is that the normalising constant $\alpha$ is not available in analytically closed form. The conditional intensity of a point process given by~\eqref{pairpdf} is given by
\begin{equation}
\lambda(u,\xx) = p(\{u\} \cup \xx)/p(\xx)
\end{equation}
taken $0/0 = 0$.
As it may be noticed, the computation of this quantity does not need $\alpha$, which allows us to work with point processes distributions. It can be interpreted as the probability of finding a new point $u$ in our point process window given the realisation $\xx$. Furthermore, it exhibits some remarkable properties. First, if the conditional intensity is bounded it implies that the model $p(\xx)$ is well defined, that is, $p(\xx)$ is integrable. Second, this boundness condition is necessary to prove desired convergence conditions for Monte Carlo dynamics used to sample from $p(\xx)$. Finally, if the point process is hereditary, that is $p(\yy) > 0 \rightarrow p(\xx) > 0$ for all $\xx \subset \yy$, then $\lambda$ uniquely determines $p$~\citep{moller2003statistical,Lies00,chiu2013stochastic,BaddEtAl16}.
For interaction free models, like the Inhomogeneous Poisson process, the conditional intensity is $\lambda(u,\xx) = b(u)$, and for pairwise interactions we have
\begin{equation}\label{lambda_pw}
\lambda(u,\xx) = b(u) \prod_{i=1}^{n(\xx)} h(u,x_i)
\end{equation}
where $h(u,x_i)$ is the corresponding interaction function at location $u$ over the whole set $\xx$. 

\subsection{Proposed Gibbs models}
In this work we will test several point process models. First, a trend function is proposed based on proximity to the filaments structure. The model assumes galaxies tend to populate the filaments creating denser regions, as found in \cite{Tempel14}. We create a spines net that traces the cosmic web structure. In addition, three different interaction models are used to model the galaxy distribution.
The interaction scale in these models is described by a radius parameter $r$, which we will vary between $0.5$ and $5\,h^{-1}\mathrm{Mpc}$, as explained in Section~\ref{methods:estim_process}.
The chosen models are the Geyer model \citep{Geyer99}, Connected Components \citep{baddeley1989nearest} and Area Interaction \citep{widom1970new, BaddLies95}. Each model creates a different interaction pattern and therefore, assumes a different galaxy interaction behaviour. 

\subsubsection*{Spines model}\label{models:spines}
In \cite{Tempel14} the authors present a cosmic web filaments catalogue obtained with the Bisous model \citep{StoiEtAl05,StoiEtAl07,stoica2010filaments}. This model detects those regions of the galaxy structure where filaments are likely to exist. These filaments are therefore identified and described as virtual spines that trace the filamentary patterns (in green in Figure~\ref{sample1}). 

The Bisous model used to detect these filaments is a marked Gibbs point process of interacting random segments forming filaments. This filamentary detector is built while no particular assumption on the galaxies distribution is made. The current work has to be seen as the dual of the previously cited one. Here, the filamentary structure of the galaxies is given and, conditionally on it, an interacting inhomogeneous Gibbs point process is fitted to the observed galaxies positions.

Within this context, let us consider the following inhomogeneous Poisson process that introduces within the intensity model a dependence of the current point position with respect to its minimum distance to the spine filaments
\begin{equation}
\label{spinepdf}
p(\xx) = \alpha \beta_1^{n(\xx)} \beta_2^{-\sum_{i=1}^{n(\xx)}d(x_i,F)}.
\end{equation}
Here $\beta_1, \beta_2 \in \mathds{R}^+$ are the model parameters, while $F$ represents the set of spines. The function $d(x,F)$ calculates the minimum distance from the point $x$ of the process to the spines pattern $F$. Whenever $\beta_2 > 1$ the points tend to aggregate around the spines (clustering behaviour), while if $\beta_2 < 1$ the points will be situated rather further away from the spine network (repulsive behaviour). The sufficient statistics vector is
\begin{equation*}
t(\xx) = (n(\xx), d_{F}(\xx))
\end{equation*}
with $d_{F}(\xx) = - \sum_{i=1}^{n(\xx)} d(x_i,F)$.
The Papangelou conditional intensity of the process writes as:
\begin{equation}\label{lambda_act}
\lambda(u,\xx) =  \beta_1 \beta_2^{-d(u,F)} \quad \text{or} \quad \log \lambda(u,\xx) =  \log \beta_1 - d(u,F) \log\beta_2 \text{.}
\end{equation}

\subsubsection*{Geyer process}
The Geyer or the saturation processes was introduced in~\cite{Geyer99} as a simple solution to produce aggregated point patterns. This model is an alternative to the one initially proposed by~\cite{KellRipl76,Stra75}. The Geyer process can be used to introduce interactions among points in the previous Spines model. In this case its probability density writes as 
\begin{equation}
\label{eq:geyer}
p(\xx) = \alpha \beta_1^{n(\xx)} \beta_2^{d_{F}(\xx)} \prod_{i=1}^{n(\xx)} \gamma^{\min(s,t(x_i|\xx))},
\end{equation}
with $t(x_i|\xx)$ the number of points $x_j \neq x_i$ in the configuration $\xx$ that are situated with a distance $r$  from the point $x_i$, and $s$ the saturation parameter. This parameter limits the number of neighbours that may influence the presence of a new point in the configuration. In this work, we will set $s=2$. Whenever $\gamma > 1$ the model tends to produce aggregated patterns, and repulsive ones if $\gamma < 1$. For $\gamma=1$, the Geyer model is equivalent with the Spines model. The presence of the saturation parameter guarantees the integrability of the model. The vector of the sufficient statistics for this process is:
\begin{equation*}
t(\xx) = \left(n(\xx), d_{F}(\xx), t_s(\xx) \right)
\end{equation*}
with $t_{s}(\xx) = \sum_{i=1}^{n(\xx)}\min(s,t(x_i|\xx)))$.

The expression of the Papangelou conditional intensity of~\eqref{eq:geyer} is:
\begin{equation}\label{geyersat}
  \lambda(u,\xx) =  \beta_1 \beta_2^{-d(u,F)} \prod_{i=1}^{n(\xx)}\gamma^{\min(s,t(u|\xx))} \prod_{i=1}^{n(\xx)}\gamma^{\min(s,t(x_i|\xx \cup \{u\})) - \min(s,t(x_i|\xx))}
\end{equation}
\begin{equation*}
\begin{split}
  \log\lambda(u,\xx) =  \log\beta_1 - d(u,F)\cdot \log \beta_2 + \log\gamma \Big( \sum_{i=1}^{n(\xx)} \min(s,t(u|\xx)) +\\ \sum_{i=1}^{n(\xx)} \min(s,t(x_i|\xx \cup \{u\})) - min(s,t(x_i|\xx)) \Big)
\end{split}
\end{equation*}
The role of this model is to test in which measure the distance based interactions (attraction or repulsion) among galaxies contribute to the galaxies distribution when merged together with the proximity of the filaments network.

\subsubsection*{Connected Components process (Concom)}
The Connected Components process is another distance based interaction point process. Let us consider the graph with vertices the points positions in the configuration $\xx$ and edges placed between those points situated at lower distance than a given radius $r$. A connected component of this graph is a set of vertices that can all be reached from one another~\cite{BaddEtAl16,baddeley1989nearest}. Identifying these components can be obtained by running the friends-of-friends algorithm \citep{feigelson2012modern} over a point process configuration. Embedded within the Spines model, its probability density function is
\begin{equation}
p(\xx) = \alpha \beta_1^{n(\xx)} \beta_2^{d_{F}(\xx)} \gamma^{v(\xx)},
\end{equation}
with $v(\xx) = n(\xx)- c(\xx)$ the difference between the number of points and the number of connected components. The sufficient statistics vector is: 
\begin{equation*}
t(\xx) = (n(\xx),d_{F}(\xx), v(\xx)).
\end{equation*}

The conditional intensity of the Connected Component process is expressed by~:
\begin{align}
\lambda(u,\xx) = \beta_1\beta_2^{-d_(u,F)}\gamma^{v(\xx \cup \{u\}) - v(\xx)}\\
\log\lambda(u,\xx) = \log\beta_1 - d(u,F) \log \beta_2 + (v(\xx \cup \{u\}) - v(\xx))\log\gamma 
\end{align}

Under the hypothesis of such model, the galaxies interactions and the induced distribution is explained, in addition to the proximity to the filaments, by the galaxies membership to the same component, regardless of its size or shape.

\subsubsection*{Area Interaction process (Area-Int)}
The area-interaction process was introduced by~\cite{BaddLies95}. Its probability density has the following expression:
\begin{equation*}
p(\xx) = \alpha \beta_1^{n(\xx)} \beta_2^{d_{F}(\xx)} \gamma^{-a(\xx)}\\
\end{equation*}
with $a(\xx)$ given by:
\begin{equation*}
a(\xx) = \frac{A(\xx)}{\big|b(o,r)\big|}, \qquad A(\xx) = \big| W \cap \bigcup_{i=1}^{n(\xx)} b(x_i,r)\big|.
\end{equation*}
Here $b(x,r)$ represents the ball centred in $x$ with radius $r$. Hence, $A(\xx)$ is the volume of the object given by the set union of the spheres of radius $r$ and that are centred in the points given by the configuration $\xx$. Dividing $A(\xx)$ by the volume of a sphere, allows to interpret the new statistic $a(\xx)$ as the number of points or spheres forming a structure with volume $A(\xx)$. This re-parametrisation is much more robust from a numerical point of view, whenever parameter estimation is considered. The sufficient statistics vector is: 
\begin{equation*}
t(\xx) = (n(\xx),d_{F}(\xx), a(\xx)).
\end{equation*}

The Papangelou conditional intensity is:
\begin{align}
\label{areaint}
  \lambda(u,\xx) = \beta_1 \beta_2^{-d(u,F)} \gamma^{a(u,\xx)}\\ 
  \log\lambda(u,\xx) = \log\beta_1 -d(u,F)\log \beta_2 + \log\gamma a(u,\xx)
\end{align}
where $a(u,\xx)$ is the relative volume of that part of the sphere of radius $r$ centred on $u$ that is not covered by the spheres of radius $r$ centred at the other points $x_i \in \xx$. Whenever $\gamma \in ]0,1[$ the volume induced by the points in the configuration $\xx$ tends to occupy the whole domain, hence the points exhibit a repulsive distribution. If $\gamma > 1$ the volume of the pattern induced by the points spatial distribution tends to reduce, hence forming clustered patterns. This model creates point patterns that in a certain sense ``compete" for the territory domain.

\section{Statistical inference based on the ABC Shadow algorithm}
\label{methods:estim}

The previously presented models exhibit different clustering mechanisms within an inhomogeneous environment. The inhomogeneity and the clustering interaction are controlled through the model parameters. All these processes are models of the exponential family that belong to the class of Gibbs point processes. Their probability density with respect the standard Poissonian reference can be written as follows~:
\begin{equation}
p(\xx|\theta) = \frac{\exp[-U(\xx|\theta)]}{c(\theta)} =  \frac{\exp \langle t(\xx),\theta \rangle)}{c(\theta)},
\label{eq:modelPattern}
\end{equation}
with $U$ the energy function, $\xx$ the point pattern, $\theta$ and $t(\xx)$, the parameter and the sufficient statistics vectors, respectively, and $c(\theta)$ the normalisation constant.

Let us consider that a point pattern $\xx$ is observed. This pattern represents galaxies positions in a compact region of our Universe $W$. Furthermore, the filamentary structure present in this region is known~\citep{Tempel14,TempEtAl14a,TempEtAl16}. Next suppose, that the point pattern is the realisation of point processes given by~\eqref{eq:modelPattern} and also that knowledge related to model parameters $\theta$ is available under the form of a prior $p(\theta)$.

Hence, statistical inference can be performed through the posterior distribution 
\begin{equation}
p(\theta|\xx) \propto p(\xx|\theta)p(\theta).
\label{eq:posteriorDensity}
\end{equation}
Performing statistical inference from the posterior distributions~\eqref{eq:posteriorDensity} is a classical problem in Bayesian analysis and, to the best of our knowledge, still a mathematical challenge. This is due to the fact that the normalisation constant $c(\theta)$ is available in analytic closed form just for a reduced class of models~\citep{BaddEtAl16,Lies00,moller2003statistical}.

The solution adopted here is to use for inference purposes the ABC Shadow algorithm~\citep{StoiEtAl17}. This algorithm outputs are approximate samples from the posterior distribution of interest. It can be applied to posterior distributions that are continuously differentiable with respect to the considered parameters. This is a rather strong hypothesis but reasonable since it is frequently met in practice.

The ABC Shadow algorithm combines two ideas. The first idea is given by the ABC principles \citep{BiauEtAl15,Blum10,GrelEtAl09,MariEtAl12}: using the Metropolis-Hasting (MH) algorithm, an auxiliary pattern is generated according to the chosen probability distribution conditioned on the current parameters $\theta$. Next, inspired by the auxiliary variable method introduced by~\citet{MollEtAl06}, the previous pattern is used to update the parameter values. Below the pseudo-code of the algorithm is given. All the technical details can be found in \cite{StoiEtAl17}.\\

\begin{algo}
\label{abcShadow}
{\bf ABC Shadow~:} Fix $\delta$ and $m$. Assume the observed pattern is $\yy$ and the current state is $\theta_0$.

\begin{enumerate}
\item Generate $\xx$ according to $p(\xx|\theta_0)$.
\item For $k=1$ to $m$ do
\begin{itemize}  
\item Generate a new candidate $\psi$ following the density $U_\delta(\theta_{k-1} \to \psi)$ defined by
\begin{equation*}
U_\delta (\theta \to \psi) = \frac{1}{V_\delta} \1_{b(\theta, \delta/2)}\{\psi\},
\label{uniformProposal}
\end{equation*}
with  $V_\delta$ the volume of the ball $b(\theta, \delta/2)$.
\item The new state $\theta_{k} = \psi$ is accepted with probability $\alpha_{s}(\theta_{k-1} \rightarrow \psi)$ given by
\begin{eqnarray}
\lefteqn{\alpha_{s}(\theta_{k-1} \rightarrow \theta_{k}) = }\nonumber \\
& = & \min\left\{1,\frac{p(\theta_{k}|\yy)}{p(\theta_{k-1}|\yy)}\times\frac{p(\xx | \theta_{k-1})\zeta(\theta_{k}) \1_{b(\theta_{k},\delta/2)}\{\theta_{k-1}\}}
{p(\xx | \theta_{k})\zeta(\theta_{k-1}) \1_{b(\theta_{k-1},\delta/2)}\{\theta_{k}\}}
\right\} \nonumber \\
& = & \min\left\{1,\frac{p(\yy|\theta_{k})p(\theta_{k})}{p(\yy|\theta_{k-1})p(\theta_{k-1})}\times\frac{p(\xx | \theta_{k-1})}
{p(\xx | \theta_{k})}
\right\}
\label{acceptance_probability_shadow}
\end{eqnarray}
otherwise $\theta_{k} = \theta_{k-1}$.
\end{itemize}
\item Return $\theta_m$.
\item If another sample is needed, go to step $1$  with $\theta_0 = \theta_n$.
\end{enumerate}
\end{algo}

\subsection{Estimation process}\label{methods:estim_process}

As explained above, we are working with a sample of 956 galaxies in a cube of side length $30 h^{-1}$ Mpc. The first step is to measure the sufficient statistics, the observed quantities that fully characterise the Gibbs point process. The sufficient statistics observed in our data set are used to estimate the different models. 

In our models the first two sufficient statistics always have the same value: $n(\xx) = 956$, the number of galaxies; and $d(\xx,F) = \sum_{i=1}^N d(x_i,F) = 1259.33$, the sum of the distances from each galaxy to the closest filament. These quantities do not depend on the interaction radius and are constant for all model evaluations. The interaction sufficient statistic, which determines if we are using a Geyer, Connected Component or Area Interaction model is radius dependent and we have to evaluate it for every radii. These statistics are presented in Table~\ref{suff}.

\begin{table}
\caption{Interaction sufficient statistics of our three models for different radii.}
\label{suff}
\begin{center}
\begin{tabular}{|c|ccc|}
\hline
$r~(h^-1 \mathrm{Mpc})$ & $t_s(\xx)$ & $v(\xx)$ & $a(\xx)$\\
\hline
0.5 & 1003 & 456 & -85.104 \\ 
1.0 & 1298 & 588 & -815.625 \\ 
1.5 & 1528 & 698 & -906.627 \\ 
2.0 & 1689 & 800 & -932.127 \\ 
2.5 & 1784 & 863 & -942.814 \\ 
3.0 & 1830 & 900 & -948.042 \\ 
3.5 & 1860 & 927 & -950.838 \\ 
4.0 & 1876 & 936 & -952.509 \\ 
4.5 & 1887 & 943 & -953.529 \\ 
5.0 & 1898 & 950 & -954.186 \\  
\hline
\end{tabular}
\end{center}
\end{table}

These quantities already give us an idea of the interaction scale in the point process. For the Geyer model, $t_s(\xx) = \sum_{i=1}^{n(\xx)} min(s,t(x_i,\xx)))$ is the number of galaxy pairs within distance $r$ thresholded by $s$ for each galaxy. As expected, this quantity grows with the radius, but after $r=2.5 h^{-1}$ Mpc, the number of pairs stabilises. Similar effects are found for the other models. This is a first indication of a Markov radius: for distances larger than this the process is approximately stationary. The same behaviour can be seen for $v(\xx)$ and $a(\xx)$.

Once the models are defined we can proceed to fit them. The ABC Shadow algorithm is run using the sufficient statistics from Table~\ref{suff} as observed data. Each time, the auxiliary variable was sampled using $100$ steps of a MH dynamics. The $\Delta$ parameter was set to (0.01, 0.01) and the algorithm was run for $200000$ iterations. The parameter estimation has been started for all parameters from the initial values $\log \beta_1 = \log \beta_2 = -6.5$ and $\log \gamma = 0.5$. 

\subsection{Asymptotic errors}

Using asymptotic normality of the Maximum Likelihood Estimation we can obtain an estimation of the standard errors on our parameters. For point processes this have been used in \cite{Geyer99,van2003candy}. For a set of parameters $\theta = (\log\beta_1, \log\beta_2, \log\gamma)$ we take the log likelihood ratio:
\begin{equation}\label{loglr}
l(\theta) = \log \frac{p_{\theta}(\xx)}{p_{\theta_0}(\xx)} = \log \frac{\alpha(\theta)}{\alpha(\theta_{0})} + t(\xx)^T(\theta - \theta_0)
\end{equation}
with respect to some reference value $\theta_0 \in (0,\infty)\times (0,\infty)$. As seen in \cite{geyer1992constrained,Geyer99}, $\alpha(\log \theta_0)/\alpha(\log \theta) = E_{\log \theta_0} \exp[ss(\xx)^T(\log \theta - \log \theta_0)]$ and the log likelihood ratio can be written as
\begin{equation}
l(\theta) = t(\xx)^T(\theta - \theta_0) - \log E_{\theta_0} \exp[t(\xx)^T(\theta - \theta_0)]
\end{equation}
from which we derive the score equations $\nabla l(\theta) = t(\xx) - E_{\theta}t(Y)$ and the Fisher information matrix $\nabla^2 l(\theta) = \Var_{\theta} t(Y)$. In summary, the maximum likelihood equations $E_{\theta}t(Y) = t(\xx)$ state that under $\hat{\theta}$, the expected values of the sufficient statistics must be equal to the observed values. Now, since the covariance matrix of $t(Y)$ is positive definite, equation~\eqref{loglr} is concave in $\theta$. Therefore, provided the score equations have a solution $\hat{\theta} \in \mathds{R}^{r}$, a unique maximum likelihood estimator exists and equals $\hat{\theta}$. Otherwise, a maximum may be found on the boundary of the parameter space.

This equation can be solved to obtain an approximation of the maximum likelihood estimation $\hat{\theta}$ using a Monte Carlo approximation \citep{geyer1992constrained,geyer1994convergence,Geyer99}. The resulting estimation is distributed as
\begin{equation}
\sqrt{m}(\theta - \theta_0) \sim N(0,I(\hat{\theta})^{-1}\Sigma   I(\hat{\theta})^{-1}) \, ,
\end{equation}
where $m$ is the number of used samples of $p_{\theta(\xx)}$. We estimate the Fisher information matrix $I(\hat{\theta}) = \Var_{\hat{\theta}}t(\xx) = -\nabla^2 l(\theta)$ with $-\nabla^2 l(\hat{\theta})$ and $\Sigma$ is the asymptotic covariance matrix of the normalised Monte Carlo score $\sqrt{m}\nabla l_m(\hat{\theta})$. This matrix is estimated as well using the generated samples:
\begin{equation}
\hat{\Sigma} = \frac{C_m}{\Big(\frac{1}{m}\sum_{i=1}^m \exp [t(Y_i)^T (\hat{\theta}_m) - \theta] \Big)^2} \, ,
\end{equation}
where $C_m$ is the empirical covariance matrix of $(t(\xx)-t(Y))e^{t(Y)^T (\hat{\theta}_m - \theta_0)}$. 

It is important to choose an initial value $\theta_0$ that is not too far from the maximum likelihood estimator. In this work we generate a large number of samples using known parameters $\theta$. Then these parameters are estimated using the resulting sufficient statistics $\theta_0$. The $m$ samples used in the Monte Carlo calculations are generated \textit{a posteriori} using the estimated $\theta$ value.

\section{Model choice}\label{choice}
Once the posterior distribution is sufficiently mapped, we can estimate the maximum a posteriori (MAP) as the best fit parameters, perform statistical inference and validate the obtained results. The used techniques are as follows.

\subsection{Residuals analysis}\label{methods:choice_residuals}

The calculation of the residuals is based on the \textit{observation minus predicted} comparison. We use the fitted model to estimate the expected number of points in a region and compare it with reality. Average zero residuals are guaranteed by construction and we can use residuals to detect over and underestimations of the model. Details can be found in \cite{RSSB:RSSB519}. The residuals definition used in this work are the \textit{raw residuals}. For an estimated model $\hat{\lambda}_{\theta}$ we have:

\begin{equation}
R(B,1,\hat{\theta}) = n(\xx \cap B) - \int_B \hat{\lambda}_{\theta} (u;\xx) \text{d}u
\end{equation}

Following the general strategy of the point processes analysis, the residuals can be evaluated as well at the empty location $u$. The $absence$ of points at these locations is also informative.

\subsubsection*{Smoothed residuals plots}

The above defined residuals can be calculated for all $u \in W$ points and therefore we can create plots to visualise the quality of our fitting. If the model is well fitted, one expects to find an uncorrelated distribution in the residual map, with no correlations between the values of the residuals and the locations of the data points. Residuals must be of low amplitude compared with the conditional surface density function and must oscillate around zero. Since we are using three dimensional point processes, we will use as integrating regions $B$ the volume covered by small cubes whose residuals will be summed along axis $Y$.

However, for the proper visualisation of the residual map, a smoothing should be done. When evaluated for a large amount of $u$ points in $W$ the map may appear too atomised, showing values close to 1 when we are close to a data point, and close to 0 for the rest. We proceed with a smoothing of the map, understanding that the shape of the structures in the pattern and its general trend are bigger than the area occupied by our numerical integration cells. A spatial distribution $\xx$ in $W$ with $N$ points can be smoothed by

\begin{equation}\label{ker}
\lambda^*(u) = \sum_{i=1}^N \kappa_{\omega}(u-x_i)e(x_i;\xx \char`\\ \{x_i\}) \text{,} \qquad u \in W
\end{equation}

where the kernel function $\kappa_{\omega}$ is the Gaussian filter for a bandwidth $\omega$. The density field $\lambda^*$ is an estimation of the local density of the distribution for every location $u$ that involves all the existing points $x_i$ in $\xx$. 

Similarly, we define the smoothed version of the estimated intensity function as

\begin{equation}\label{dag}
\lambda^{\dag}(u) = \int_{W} \kappa_{\omega}(u-v)e(u;\xx) \hat{\lambda}_{\theta}(v) \text{d}v
\end{equation}

The final smoothed residuals are

\begin{equation}\label{eq:su}
s(u) = \lambda^*(u) - \lambda^{\dag}(u)
\end{equation}

The computation of these residuals involves several details that can be fully consulted in \cite{RSSB:RSSB519}. The smoothed residuals are integrated in a weighted calculation with edge corrections. It is important to note that the comparison between $\lambda^*$ and $\lambda^{\dag}$ is not direct since $\lambda^*$ is the data density field and $\lambda^{\dag}$ is the conditional probability density function. $\lambda^{\dag}$ might be interpreted as the probability of finding a point in a location $u$ given the rest of the data. 

\subsection{Q-Q plots}\label{methods:choice_qqplots}

The quantile-quantile plot of residuals compares the residuals obtained from the analysed data set and those predicted by the theoretical distribution. \cite{RSSB:RSSB519} recommend this test as a useful diagnostic for assessing the modelling of the interaction part of the model.

The residual values obtained with function $s(u)$ are ordered and compared with those from the model. Using MH and the estimated parameters $\hat{\theta}$ from our original data set we generate several simulated samples (typically a hundred). If the model is correct and $\hat{\theta}$ are well estimated, those simulations should be realisations of the same stochastic process. Using the best fit parameters we calculate the $s(u)$ residuals as well. When ordering these values we should obtain similar results to those of the original data set. A confidence region is created by selecting the 2.5\% and the 97.5\% quantiles of the simulated residuals.

Then we plot the data quantile ($Y$ axis) against the averaged simulations quantiles ($X$ axis) together with the confidence region limits (in red). Interpretation of this plot is highly informative but complex. Details can be found on \cite{BaddEtAl16}.

\section{Results}\label{results}

In this section we present the different results obtained with our data set, for the best fit estimated model parameters. For better understanding of our results we estimate the four proposed models: the spline model (i.e., an Inhomogeneous Poisson process with no interaction) and the trend plus interaction models: Geyer, Connected Components and Area Interaction.

As explained in section~\ref{methods:estim_process}, we proceed to estimate the interaction models for radii $r = 0.5 \,h^{-1}$ Mpc to $5 \,h^{-1}$ Mpc with steps of $0.5$. The obtained parameters are presented in Tables~\ref{geyer_pars}--\ref{areaint_pars} and Figures~\ref{geyer_box}--\ref{areaint_box}, and described below.

\subsection*{Inhomogeneous Poisson process}

This first model uses only the spline intensity function as introduced in equation~\eqref{spinepdf} with constant interaction function $h(x_i,x_j)=1$.

We have two sufficient statistics $ss(\xx) = (956,1259.33)$ and no dependency on the radius, hence we only have to estimate the model once. For our galaxy sample the resulting parameters are $\log \beta_1 = -0.34 \pm 0.05$ and $\log \beta_2 = 1.32 \pm 0.03$. The splines intensity parameter, $\beta_2$ is positive and dominant, which creates a pattern where most of the galaxies are distributed along the splines. This is in agreement with observations, where galaxies are mostly concentrated around filaments and only a few are scattered outside. The later are generated in our model by the much smaller parameter $\beta_1$, i.e., with a weaker impact.

\subsection*{Geyer model}

It is known that galaxy interactions inside clusters or filaments are far from linear. This justifies the inclusion of a new component in our model, an interaction component. We start with the Geyer model, where interactions between points are determined by distance separation. With this model, galaxies are understood to cluster at distances below an interaction radius, while showing an independent Poisson-like distribution for greater distances. To prevent a non integrable accumulation of galaxies the saturation limit is fixed at $s=2$.

\begin{figure}
  \centering
\includegraphics[width=0.33\textwidth]{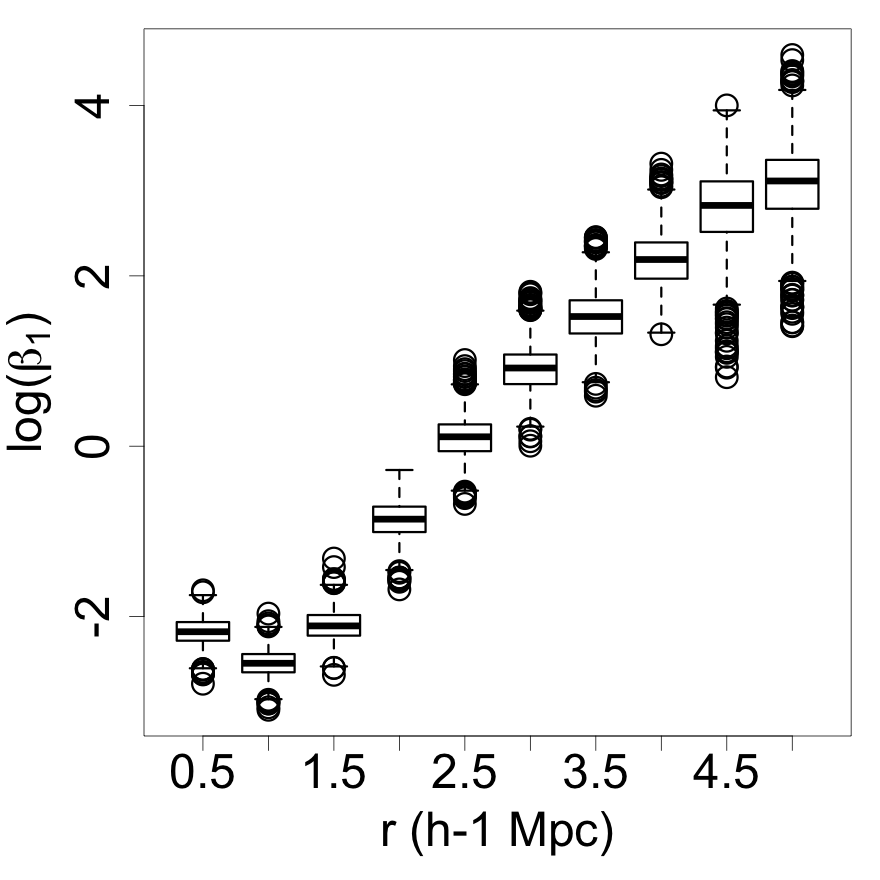}
\includegraphics[width=0.33\textwidth]{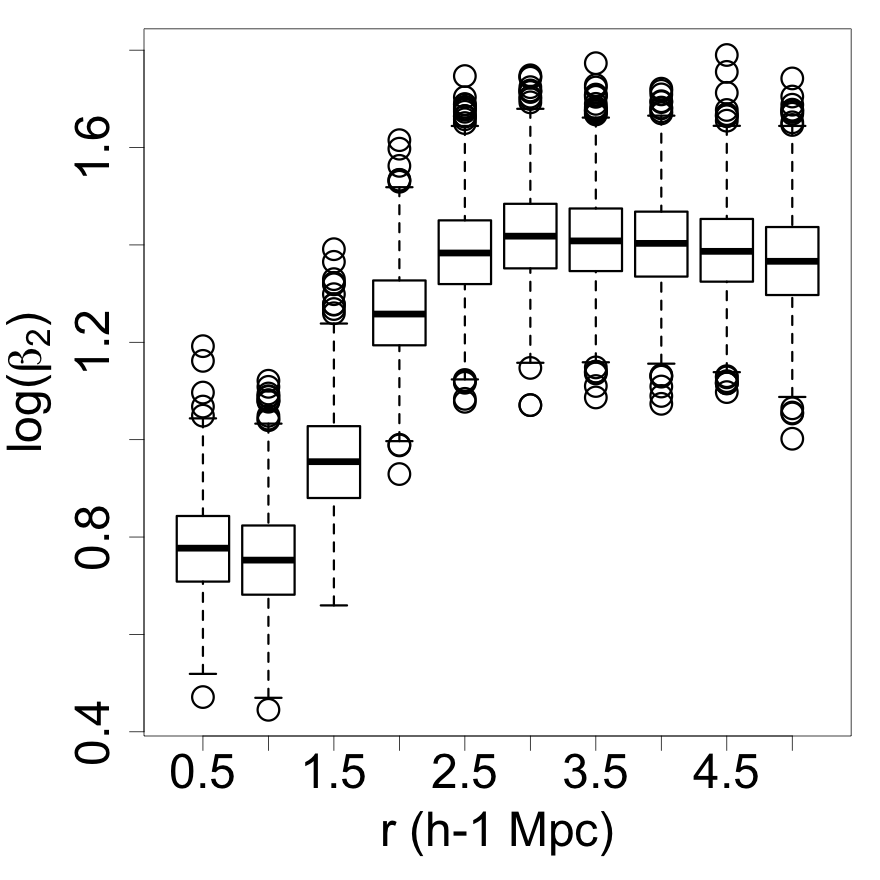}
\includegraphics[width=0.33\textwidth]{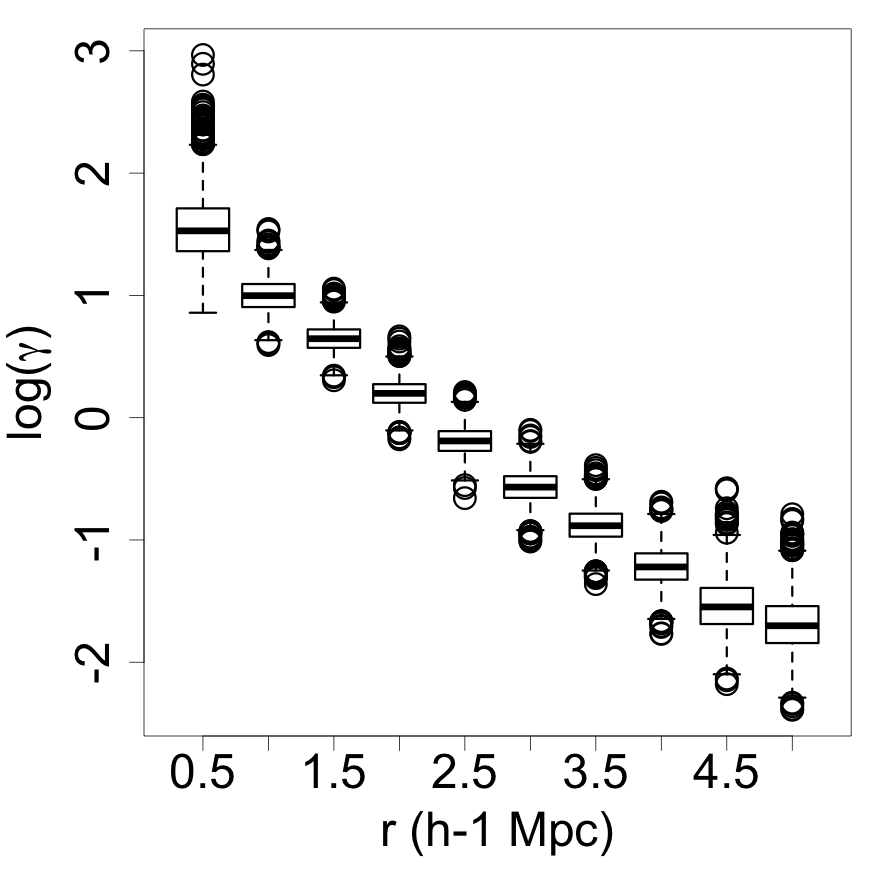}
\caption{Evolution of box plot distribution for Geyer model parameters with different interaction radii. From top to bottom: parameters $\log \beta_1$, $\log \beta_2$ and $\log \gamma$}\label{geyer_box}
\end{figure}

The estimated parameters can be seen in Table~\ref{geyer_pars} and their box plot distributions are in Figure~\ref{geyer_box}. The results indicate a negative correlation between the intensity parameters ($\log \beta_1$ and $\log \beta_2$) and $\log \gamma$ (level of interaction) for distances below $2.5 h^{-1}$ Mpc. This is expected since high values of $\log \gamma$ contribute to aggregate more points in denser regions and, as a consequence, the other parameters need to be smaller to compensate the effect and maintain the total number of points $N$. In addition, $\log \gamma$ takes value 0 for interaction distance $r$ around $2.5 h^{-1}$ Mpc, meaning that the clustered structures found by the Geyer model in our data set have sizes smaller than this range, this is in agreement with the Markov radius found in section~\ref{methods:estim_process}. Interestingly, parameter $\beta_2$ reaches a constant value for distances greater or equal than this range. In \cite{Tempel14}, the authors use radius $3 h^{-1}$ to create the filaments used in this work. The thickness of filaments in the cosmic web is thought to be of that scale, as supported by \cite{pimbblet2004intracluster,Smith2012multiscale}. As a conclusion, the Geyer model produce results compatible with our understanding of the cosmic web distribution.

\begin{table}
\caption{Best fit parameters for Geyer model with different interaction radii.}
\label{geyer_pars}
\begin{center}
\begin{tabular}{|c|ccc|}
\hline
 & \multicolumn{3}{c|}{Geyer model}\\
$r~(h^-1 \mathrm{Mpc})$ & $\log \beta_1$ & $\log \beta_2$ & $\log \gamma$ \\
\hline
0.5 & $-2.22\pm0.08$ & $0.77\pm0.03$ & $1.48\pm0.03$ \\
1 & $-2.53\pm0.1$ & $0.74\pm0.04$ & $1.01\pm0.04$ \\
1.5 & $-2.13\pm0.14$ & $0.96\pm0.04$ & $0.64\pm0.05$ \\
2 & $-0.85\pm0.18$ & $1.25\pm0.05$ & $0.2\pm0.07$ \\
2.5 & $0.16\pm0.19$ & $1.37\pm0.04$ & $-0.2\pm0.08$ \\
3 & $0.9\pm0.2$ & $1.43\pm0.04$ & $-0.58\pm0.09$ \\
3.5 & $1.5\pm0.3$ & $1.39\pm0.04$ & $-0.89\pm0.12$ \\
4 & $2.2\pm0.3$ & $1.41\pm0.05$ & $-1.24\pm0.13$ \\
4.5 & $2.9\pm0.3$ & $1.37\pm0.05$ & $-1.58\pm0.12$ \\
5 & $3.2\pm0.4$ & $1.36\pm0.04$ & $-1.8\pm0.2$ \\
\hline
\end{tabular}
\end{center}
\end{table}

\subsection*{Connected Components model}

\begin{figure}
  \centering
\includegraphics[width=0.33\textwidth]{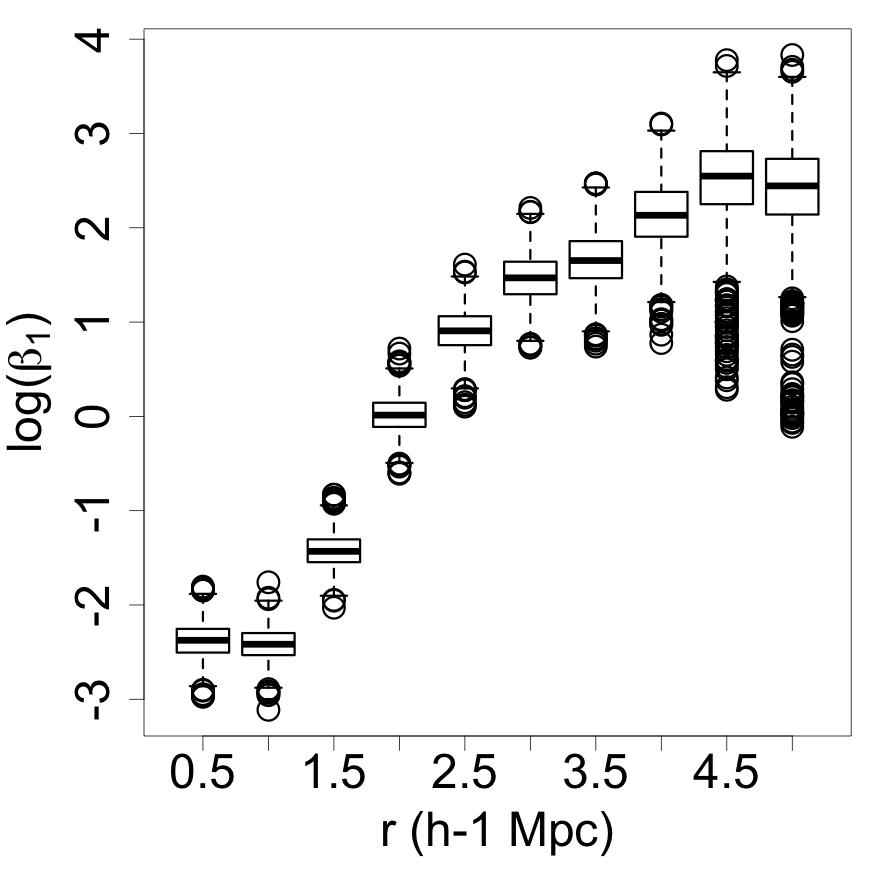}
\includegraphics[width=0.33\textwidth]{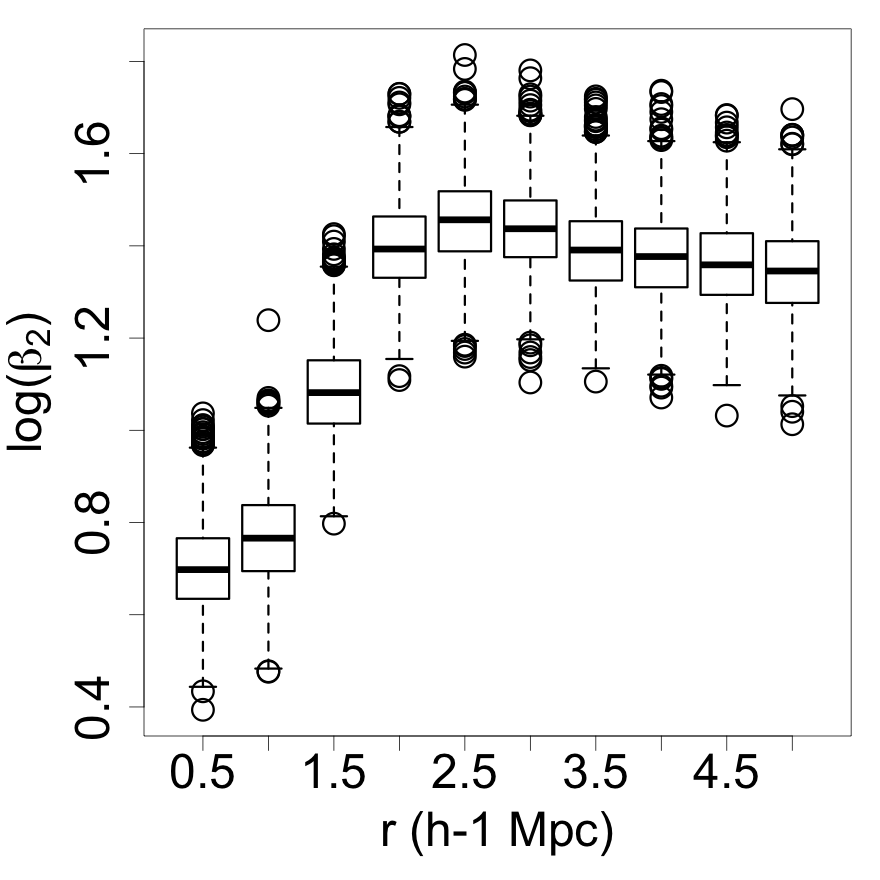}
\includegraphics[width=0.33\textwidth]{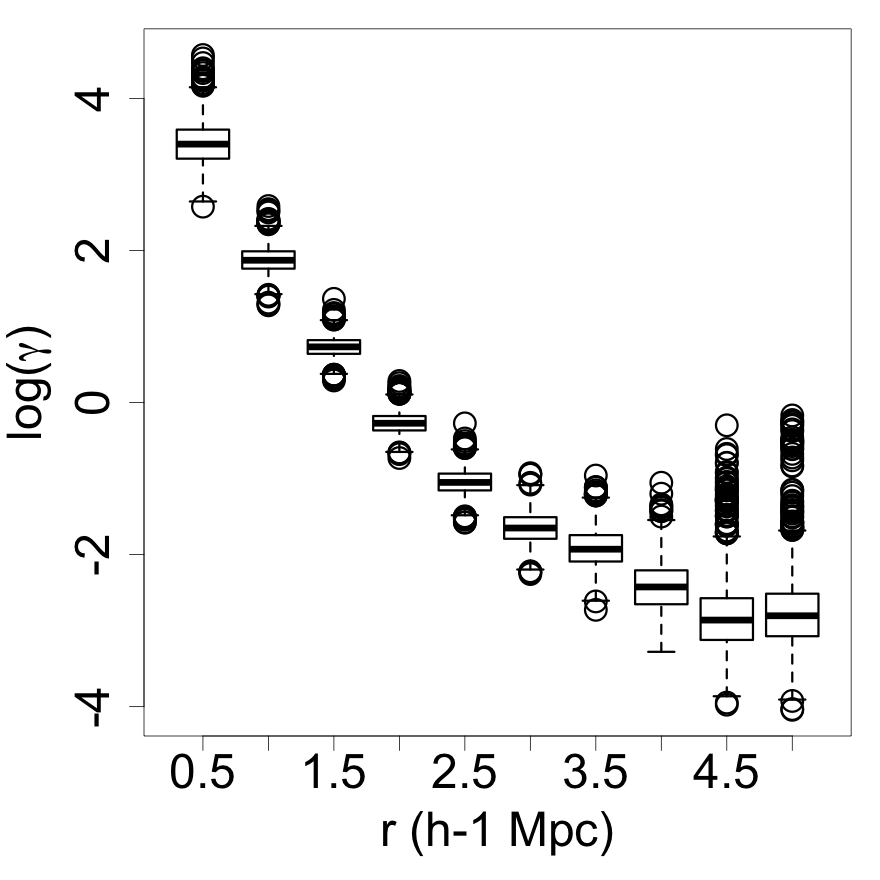}
\caption{Evolution of box plot distribution for Connected Components model parameters with different interaction radii. From top to bottom: parameters $\log \beta_1$, $\log \beta_2$ and $\log \gamma$}\label{concom_box}
\end{figure}

The best fit parameters obtained with the Connected Components model are shown in Table~\ref{concom_pars} and Figure~\ref{concom_box}. They have a similar interpretation to those obtained with the Geyer model. Both models are made to increase the level aggregation by increasing the number of points inside an overdensity (clusters grow in density, not in size). When $\gamma > 1$, this model penalises the creation of independent components, this is, isolated groups of galaxies. Given a cluster in a Connected Components process, a hypothetical new point inside that cluster has higher probabilities of belonging to the process than an isolated one. This creates the same anticorrelation effect as in the Geyer model between the intensity parameters and $\log \gamma$ at transition distance $2.5-3 h^{-1}$ Mpc and we interpret them the same way. Now clusters are defined as points belonging to the same component, as defined by the friends-of-friends algorithm. 

\begin{table}
\caption{Best fit parameters for Connected Components model with different interaction radii.}
\label{concom_pars}
\begin{center}
\begin{tabular}{|c|ccc|}
\hline
 & \multicolumn{3}{c|}{Connected Components model}\\
$r~(h^-1 \mathrm{Mpc})$ & $\log \beta_1$ & $\log \beta_2$ & $\log \gamma$ \\
\hline
0.5 & $-2.34\pm0.02$ & $0.69\pm0.03$ & $3.43\pm0.07$ \\
1 & $-2.45\pm0.05$ & $0.77\pm0.04$ & $1.83\pm0.08$ \\
1.5 & $-1.45\pm0.07$ & $1.06\pm0.05$ & $0.74\pm0.09$ \\
2 & $0.01\pm0.07$ & $1.38\pm0.05$ & $-0.28\pm0.1$ \\
2.5 & $0.84\pm0.07$ & $1.46\pm0.04$ & $-1.08\pm0.13$ \\
3 & $1.46\pm0.06$ & $1.45\pm0.04$ & $-1.65\pm0.16$ \\
3.5 & $1.56\pm0.07$ & $1.41\pm0.05$ & $-2\pm0.19$ \\
4 & $2.09\pm0.07$ & $1.39\pm0.04$ & $-2.4\pm0.2$ \\
4.5 & $2.6\pm0.07$ & $1.34\pm0.05$ & $-3\pm0.3$ \\
5 & $2.44\pm0.07$ & $1.36\pm0.05$ & $-2.9\pm0.4$ \\
\hline
\end{tabular}
\end{center}
\end{table}

\subsection*{Area Interaction model}

The Area Interaction model has a different functioning. The cluster is defined in terms of volume domination or occupancy. Structures grow in size more than intensity, adding new points to the borders. This way the cluster grows in size around the filaments with an interaction level determined by parameter $\log \gamma$. 

\begin{figure}
  \centering
\includegraphics[width=0.33\textwidth]{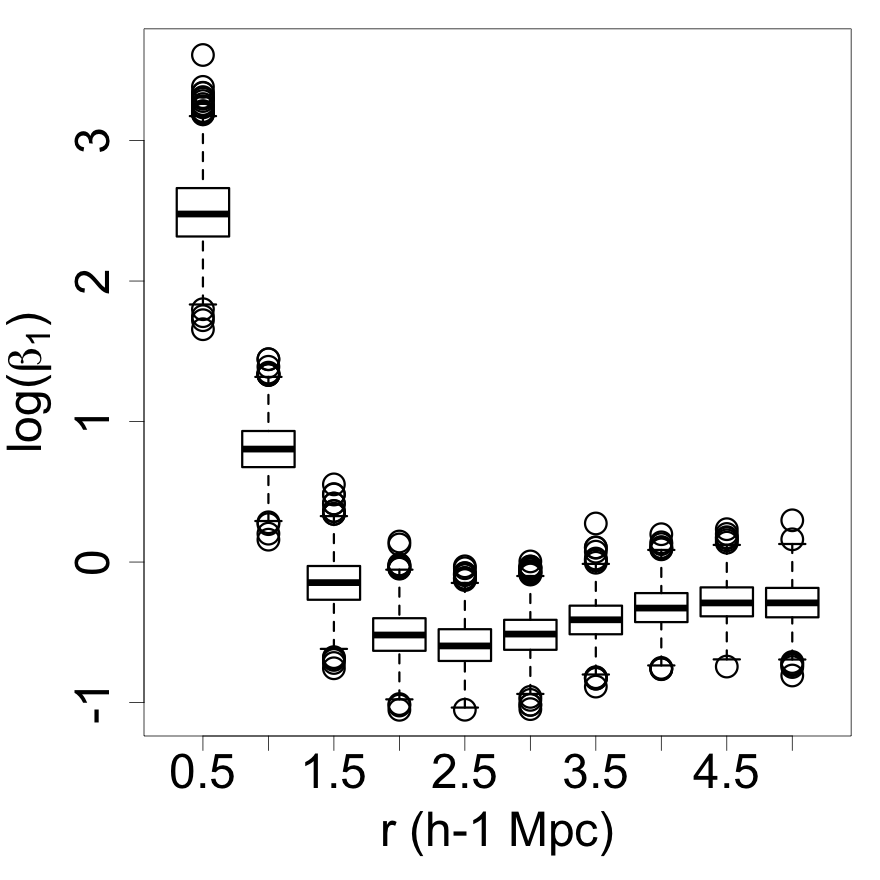}
\includegraphics[width=0.33\textwidth]{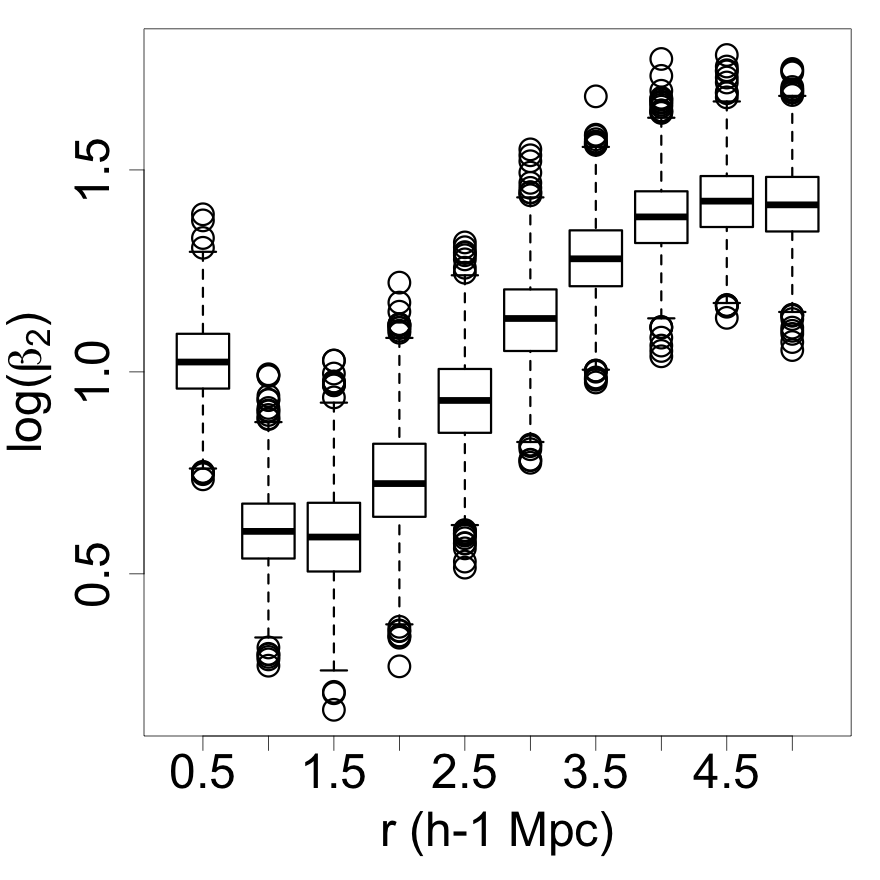}
\includegraphics[width=0.33\textwidth]{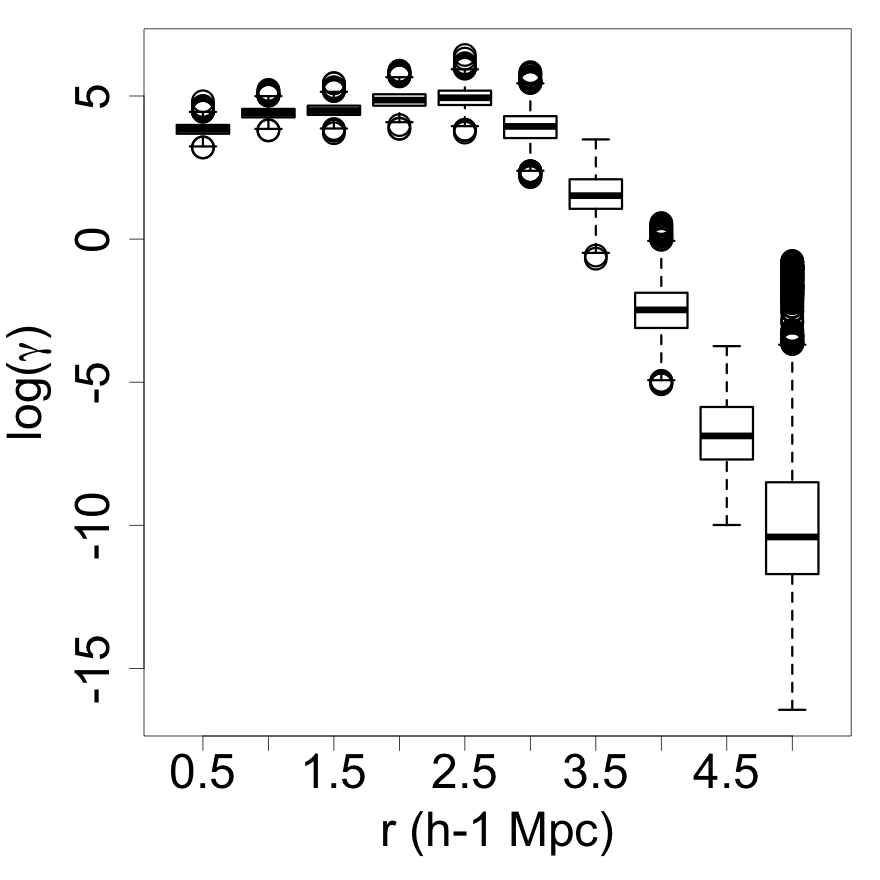}
\caption{Evolution of box plot distribution for Area Interaction model parameters with different interaction radii. From top to bottom: parameters $\log \beta_1$, $\log \beta_2$ and $\log \gamma$}\label{areaint_box}
\end{figure}

The results, shown in Table~\ref{areaint_pars} and Figure~\ref{areaint_box}, show a different behaviour when compared with the previous models. At small scales (below $3 h^{-1}$ Mpc) the interaction parameter remains approximately constant: $\log \gamma$ is significantly different from zero, proving the presence of a clustering process, but no dependence with radius is found. However, a variation is found for the intensity parameters: while $\log \beta_1$ decreases with radius from $0.5$ to $3 h^{-1}$ Mpc, the $\log \beta_2$ parameter shows a minimum between $r=1$ and $1.5 h^{-1}$ Mpc. Given a constant level of aggregation ($\log \gamma \simeq 4.3$), we can imagine galaxy clustering determined by this parameter in a stationary process. At range $r=0.5 h^{-1}$ Mpc, both $\log \beta_1$ and $\log \beta_2$ show local maximums. As this interaction range grows, the intensity parameters decrease, as the size of clusters and filaments grow and more galaxies are aggregated. Remember that the first two sufficient statistics remain constant. This process continues until a distance of $1$ or $1.5\,h^{-1} \mathrm{Mpc}$ is reached, where $\log \beta_2$ starts growing. This trend shows a convex like figure, with distance $1 h^{-1}$ Mpc as a critical point for $\log \beta_2$.

After $3 h^{-1}$ Mpc, the opposite effect is found, with $\log \beta_1$ nearly constant and $\log \gamma$ rapidly decreasing, taking negative values (i.e., repulsive pattern). The $\log \beta_2$ takes instead bigger values again to populate the process. As with the previous models, we will not consider repulsive patterns as good descriptors of our galaxy sample.

\begin{table}
\caption{Best fit parameters for Area Interaction model with different interaction radii.}
\label{areaint_pars}
\begin{center}
\begin{tabular}{|c|ccc|}
\hline
 & \multicolumn{3}{c|}{Area Interaction model}\\
$r~(h^-1 \mathrm{Mpc})$ & $\log \beta_1$ & $\log \beta_2$ & $\log \gamma$ \\
\hline
0.5 & $2.45\pm0.04$ & $1.02\pm0.03$ & $3.82\pm0.13$ \\
1 & $0.84\pm0.03$ & $0.59\pm0.03$ & $4.371\pm0.017$ \\
1.5 & $-0.12\pm0.04$ & $0.58\pm0.03$ & $4.473\pm0.009$ \\
2 & $-0.56\pm0.05$ & $0.69\pm0.04$ & $4.863\pm0.005$ \\
2.5 & $-0.63\pm0.07$ & $0.95\pm0.05$ & $4.969\pm0.004$ \\
3 & $-0.48\pm0.07$ & $1.15\pm0.06$ & $4.013\pm0.006$ \\
3.5 & $-0.39\pm0.07$ & $1.28\pm0.05$ & $1.315\pm0.005$ \\
4 & $-0.36\pm0.07$ & $1.39\pm0.05$ & $-2.013\pm0.005$ \\
4.5 & $-0.31\pm0.07$ & $1.43\pm0.05$ & $-7.131\pm0.004$ \\
5 & $-0.3\pm0.3$ & $1.4\pm0.3$ & $-11.345\pm0.011$ \\
\hline
\end{tabular}
\end{center}
\end{table}

\subsection{Model quality}\label{results:quality}

\begin{figure*}
  \centering
\includegraphics[width=0.32\textwidth]{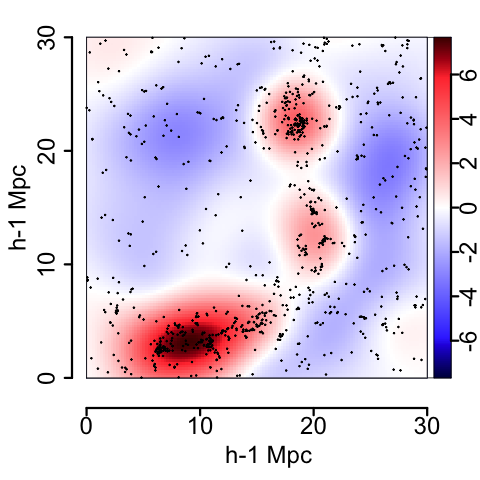}
\includegraphics[width=0.32\textwidth]{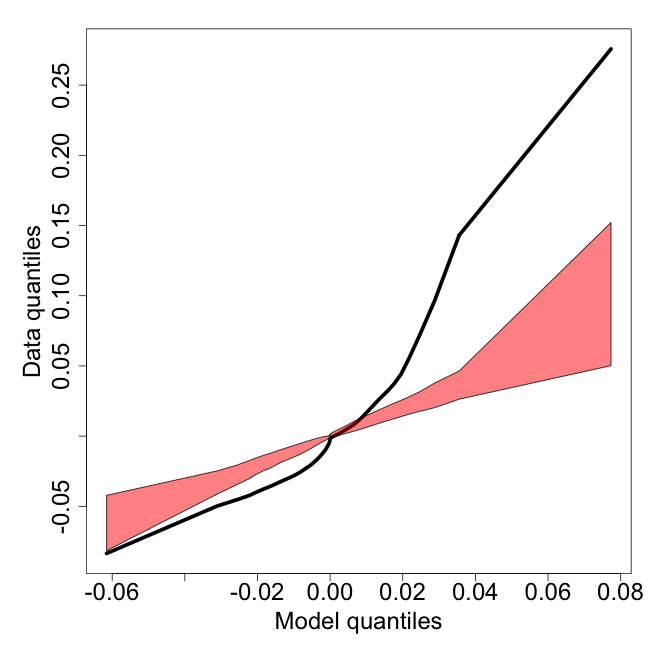}\\
\includegraphics[width=0.32\textwidth]{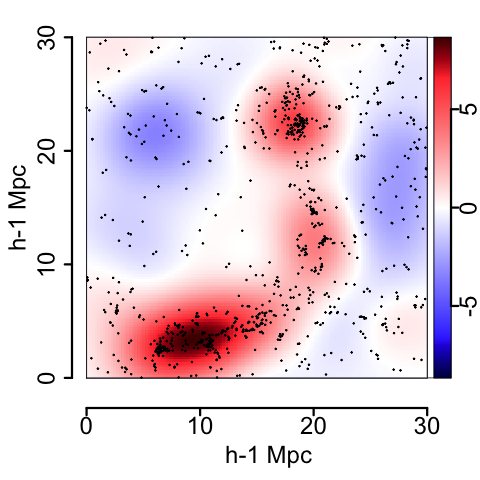}
\includegraphics[width=0.32\textwidth]{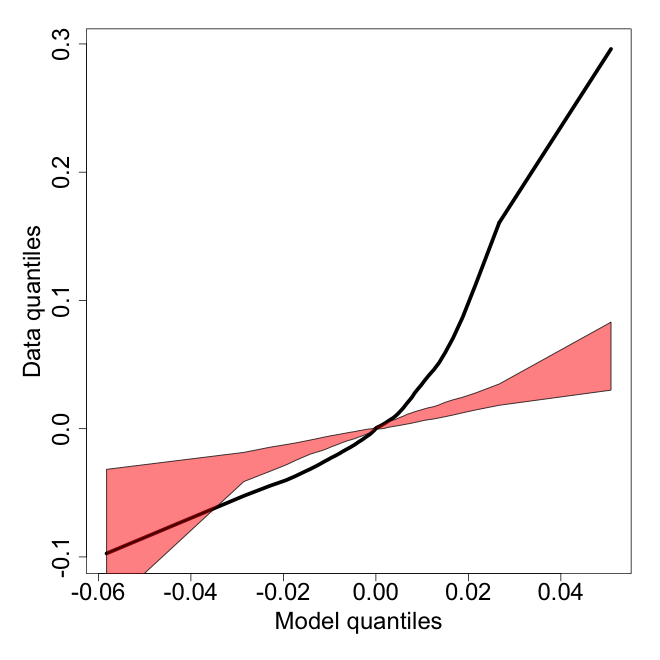}\\
\includegraphics[width=0.32\textwidth]{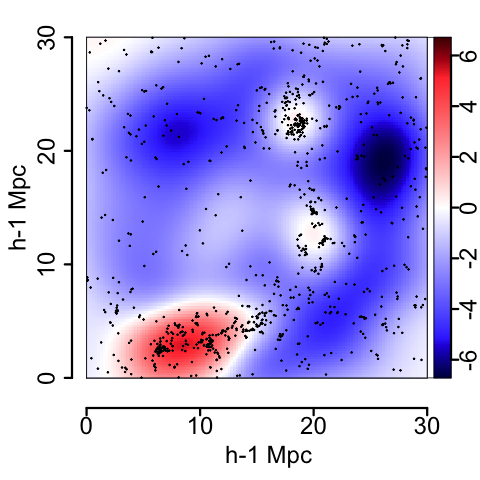}
\includegraphics[width=0.32\textwidth]{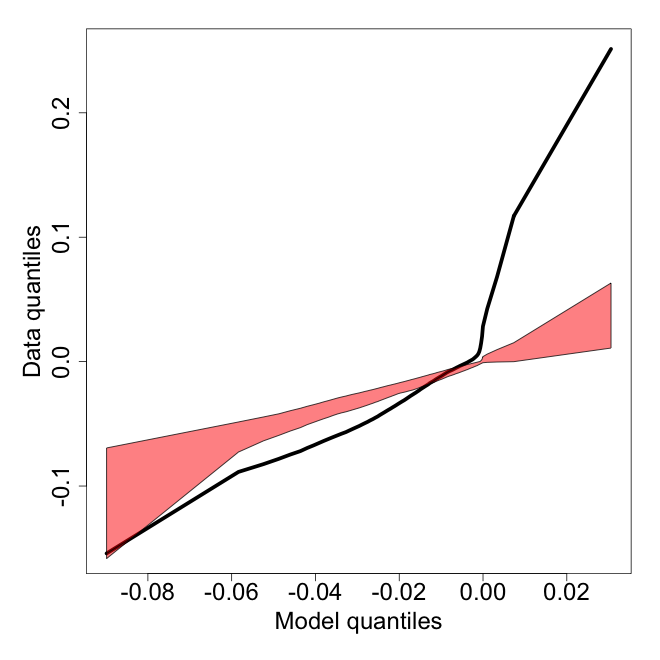}\\
\includegraphics[width=0.32\textwidth]{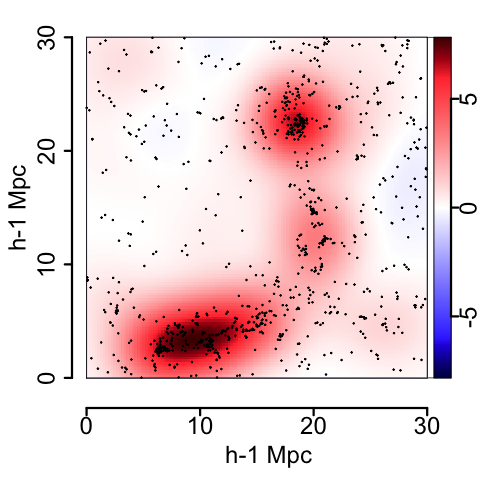}
\includegraphics[width=0.32\textwidth]{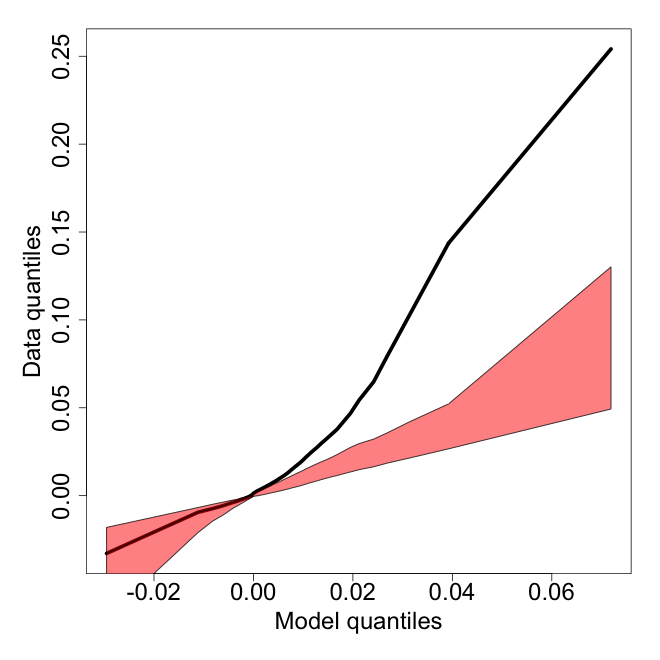}\\
\caption{Left column: residuals $s(u)$ for $r=1\,h^{-1}$ Mpc smoothed with bandwidth $\omega = 2.5 \, h^{-1} \mathrm{Mpc}$ and summed along axis Y. Right column: Q-Q plots with $r=1\,h^{-1}$ Mpc. Top to bottom: Inhomogeneous Poisson, Geyer, Connected Components and Area Interaction models.}
\label{all_res}
\end{figure*}

Regarding the quality of the models, we present in Figure~\ref{all_res} the smoothed residuals and the Q-Q plots (see section~\ref{methods:choice_residuals}). In the residual maps, we show axes X and Z of the galaxy sample with residuals summed over axis Y. This allows us to detect any fitting issue locally, and clearly identify which structures are worse fitted by the model. For comparison, all residual functions (left panels in Figure~\ref{all_res}) are graded with the same colour scale. Darker colours show a worse fitting, while fainter colours are a sign of small residuals. 

Right to these residuals we show the Q-Q plot calculations. For the interaction models we calculated the residuals with multiple radii and chose $r = 1\, h^{-1} \mathrm{Mpc}$ as the best fitting radius, therefore, we show the results with this radius. In these plots we compare the distribution of the residuals after estimating a model. If a model is correctly describing a data set, the amplitude and distribution of the residuals should be similar when estimating the real data set or a realisation of the model. The red lines in the Q-Q plots represent the 95\% confidence region for our model realisations, while the black line corresponds to the residuals obtained with the model. 
As all residuals are 0 on average, all curves cross at (0,0).
However, for other values the black curve is significantly outside the confidence region. Again, the Area Interaction model shows the best agreement, with small underdensities (negative values of the curves) and values inside the confidence region. In our opinion, this proves a satisfactory modelling of the data set for regions outside the clusters, where residuals have smaller vales (white or bluish in the figure). These regions correspond to the large scales, where galaxy distances are higher. 
However, for clusters and other overdensities, all the models show a departure from the data, with an underestimation of the real intensity of points. As said, this is mainly due to the presence of dense structures at ranges below the $1\, h^{-1}$ Mpc.  

\section{Conclusions}\label{conclusions}

This work proposes a new methodological approach to fit Gibbs point processes to characterise galaxy distributions.
To the best of our best knowledge, this is the first time that interacting inhomogeneous Gibbs point processes are fitted to this type of data conditionally on the filaments pattern (see~\citealt{lidayi21} for a similar work on star clusters). 
The work is important because it completes the existing work on filaments detection in the galaxy distribution. 
This opens the question related to the possibility of building new methodologies for analysing cosmological data that  are able to simultaneously detect patterns and estimate their characteristics.
Some necessary steps are to be mentioned before tackling such an ambitious perspective. 
First, the improvement of the model fitting by considering multiple interactions. Second, the improvement of the computational efficiency in order to be able to test this method on larger data sets by considering parallelisation of the existing algorithms.

In this work we tried to obtain a satisfactory model for the galaxy distribution. This was done on a single data set of 956 galaxies from the SDSS DR8, and three different intensity plus interaction Gibbs models. We have proved that the intensity component, based on the proximity of galaxies to the filaments is necessary to correctly describe the large scale structure of the galaxy distribution.
However, this intensity-only model lacks an interaction component, necessary to describe the way galaxies condition their distribution to other neighbours. This is a distance based interaction, which \cite{Tempel14} found around $1.5 h^{-1}$ Mpc for filaments. For this interaction radius, the estimated model parameters are significantly greater than zero for all models, indicating the necessary contribution of each component to the model and a clustering pattern.

Another distance of interest is what we call the transition distance, which we find around $3 h^{-1}$ Mpc. For interaction ranges greater than distance, the sufficient statistics of our models stabilise at constant values, showing no dependence with the interaction range.

In \cite{2018A&A...618A..81T} the authors use a combination of the Strauss and Area Interaction model to detect clusters of galaxies and other structures. We stick to this model, Area Interaction, and discuss the possible cosmological implications of such a model.

Area Interaction processes have to be understood as a volume domination process: for each point or galaxy in the process, a sphere of radius $r$ centred on it is considered. Hence, when a new point is considered, the model treats differently locations already covered by spheres than those in not yet covered volumes. 
For clustering processes this creates a pattern where overdensities, such as clusters and filaments, tend to grow by accreting new galaxies to their borders, where uncovered volume can be covered. This mechanism defines clusters in Area Interaction processes as structures growing in size more than density. This is therefore the characterisation obtained for our structures in this analysis.

However, this model is conditioned by the chosen interaction radius. This range works as a rule of measure, and no structures smaller than this distance can be correctly described by the model. Previous works \cite{Tempel14} support the value $r=1.5 h^{-1}$ Mpc for an interaction radius. Therefore, dense clusters and groups of galaxies with pairs of galaxies separated by smaller distances would be systematically underestimated by the model.

The definition of the Area Interaction model as based in the volume dominance of a central galaxy makes it comparable to the Halo Occupation Distribution \citep[HOD,][]{ 2002PhR...372....1C, 2002ApJ...575..587B, 2003ApJ...593....1B, 2004ApJ...609...35K}. This model defines the probability distribution for a dark matter halo of mass $M$ to host a number $N$ of galaxies (which could be associated in this context to subhalos). 
A possible approach to this model distinguishes two separated galaxy populations: central halo galaxies (zero or one per halo) and satellite galaxies within halos, which are separately modeled by the HOD \citep{2002MNRAS.335..311G}. In its most simple formalism of this distribution, central galaxies can be modeled as a step function.
Then, halos with masses below a threshold $M_{min}$ are expected to contain zero galaxies, while those with higher masses will contain one central galaxy on average. 
Satellite galaxies are then added with its number having a power-law dependence on the halo mass $M$, and their spatial distribution inside the halo following its matter density profile.
Given that within the halo model there is a one-to-one relation between the mass of a halo and its corresponding virial radius, this dependence of $N$ on $M$ could be also interpreted as a dependence on the size of the halo.
In addition, the HOD proves that the formation of halos and their subsequent merging and dynamical evolution are the main processes shaping galaxy clustering.

The Area Interaction model partially reproduces the central halo galaxies distribution, creating a sphere around every galaxy and merging them to define covered locations. Although no mass threshold is used in our work and each galaxy holds its own halo, this partial similarity with the HOD formalism encourages our preference for the Area Interaction model. The comparison with the HOD suggests new improvements for future works, such as using marked point processes to distinguish between galaxies of different masses or combining different models for central and satellite galaxies. Additionally, for a future work, we aim to improve our model with a combination of an Area Interaction process with a Strauss process, as used by \cite{2018A&A...618A..81T}. This model is expected to correct the underestimation at small scales.

\section*{Acknowledgements}
This work has been funded by the project PID2019-109592GB-100 from the Spanish Ministerio de Ciencia e Innovaci\'on, by the Project of excellence Prometeo/2020/085 from the Conselleria d'Innovaci\'o, Universitats, Ci\`encia i Societat Digital de la Generalitat Valenciana, and by the Acci\'on Especial UV-INV-AE19-1199364 from the Vicerrectorado de Investigaci\'on de la Universitat de Val\`encia. 

This work uses data derived from the SDSS. Funding for the SDSS and SDSS-II has been provided by the Alfred P. Sloan Foundation, the Participating Institutions, the National Science Foundation, the U.S. Department of Energy, the National Aeronautics and Space Administration, the Japanese Monbukagakusho, the Max Planck Society, and the Higher Education Funding Council for England. The SDSS Web Site is http://www.sdss.org/. The SDSS is managed by the Astrophysical Research Consortium for the Participating Institutions. The Participating Institutions are the American Museum of Natural History, Astrophysical Institute Potsdam, University of Basel, University of Cambridge, Case Western Reserve University, University of Chicago, Drexel University, Fermilab, the Institute for Advanced Study, the Japan Participation Group, Johns Hopkins University, the Joint Institute for Nuclear Astrophysics, the Kavli Institute for Particle Astrophysics and Cosmology, the Korean Scientist Group, the Chinese Academy of Sciences (LAMOST), Los Alamos National Laboratory, the Max-Planck-Institute for Astronomy (MPIA), the Max-Planck-Institute for Astrophysics (MPA), New Mexico State University, Ohio State University, University of Pittsburgh, University of Portsmouth, Princeton University, the United States Naval Observatory, and the University of Washington.  

\section*{Data availability}

The data underlying this article were obtained by \citet{Tempel14} 
and were accessed from the Strasbourg astronomical Data Center (CDS) at 
\url{https://cdsarc.u-strasbg.fr/viz-bin/cat/J/MNRAS/438/3465}. The derived 
data generated in this research will be shared on reasonable request to 
the corresponding author.





\bibliographystyle{mnras}
\bibliography{bibliography}








\bsp	
\label{lastpage}
\end{document}